\documentclass[11pt,a4paper]{article}
\usepackage[marginratio=1:1,height=600pt,width=460pt,tmargin=100pt]{geometry}
\usepackage[utf8]{inputenc}
\usepackage[parfill]{parskip}    
\usepackage{url}            
\usepackage{booktabs}       
\usepackage{amsfonts}       
\usepackage{nicefrac}       
\usepackage{microtype}      
\usepackage[dvipsnames]{xcolor}
\usepackage{graphicx}
\usepackage{amsmath}
\usepackage{amssymb}
\graphicspath{{./pics/}}
\usepackage{subcaption}
\usepackage{amsthm}
\usepackage{amscd}
\usepackage{mathrsfs}
\usepackage{bm}
\usepackage[linesnumbered,ruled,vlined]{algorithm2e}
\usepackage{multirow}
\usepackage{mathtools}
\usepackage{hyperref}
\usepackage{url}
\usepackage{enumitem}
\usepackage{authblk}
\usepackage{overpic}
\usepackage[normalem]{ulem}
\mathtoolsset{showonlyrefs}  
\usepackage{natbib}
\newtheorem{example}{Example}
\usepackage{verbatim}

\allowdisplaybreaks

\newcommand{\R}{\mathbb{R}}

\newcommand{\bfx}{\mathbf{x}}
\newcommand{\bx}{\mathbf{x}}
\newcommand{\bX}{\mathbf{X}}
\newcommand{\bA}{\mathbf{A}}

\newtheorem{remark}{Remark}

\newtheorem*{thm*}{Theorem}

\newtheorem*{lemma*}{Lemma}

\usepackage{array}
\newcolumntype{C}[1]{>{\centering\let\newline\\\arraybackslash\hspace{0pt}}m{#1}}

\SetKwInput{KwInput}{Required}                
\SetKwInput{KwOutput}{Output}              

\title{Robust Moment Identification for Nonlinear PDEs via a Neural ODE
Approach}
\author[1]{Shaoxuan Chen\footnote{Email: shaoxuanchen@umass.edu.}}
\author[1]{Su Yang}
\author[1]{Panayotis G.~Kevrekidis}
\author[2]{Wei Zhu}

\affil[1]{Department of Mathematics and
Statistics, University of Massachusetts
Amherst, Amherst, MA 01003-4515, USA}
\affil[2]{School of Mathematics, Georgia Institute of Technology, Atlanta, GA 30332, USA
}

\begin{document}

\maketitle

\begin{abstract}
We propose a data-driven framework for learning reduced-order moment dynamics from PDE-governed systems using Neural ODEs. In contrast to derivative-based methods like SINDy, which necessitate densely sampled data and are sensitive to noise, our approach 
based on Neural ODEs directly models moment trajectories, enabling robust learning from sparse and potentially irregular time series.
Using as an application platform the nonlinear Schr\"{o}dinger equation, the framework accurately recovers governing moment dynamics when closure is available, even with limited and irregular observations. For systems without analytical closure, we introduce a data-driven coordinate transformation strategy based on Stiefel manifold optimization, enabling the discovery of low-dimensional representations in which the moment dynamics become closed, facilitating interpretable and reliable modeling. We also
explore cases where a closure model is not known, such as a 
Fisher–KPP reaction–diffusion system. Here we demonstrate that Neural ODEs can still effectively approximate the unclosed moment dynamics and achieve superior extrapolation accuracy compared to 
physical-expert-derived ODE models. This advantage remains robust even under sparse and irregular sampling, highlighting the method’s robustness in data-limited settings.
Our results highlight the Neural ODE framework as a powerful and flexible tool for learning interpretable, low-dimensional moment dynamics in complex PDE-governed systems.

\end{abstract}

{
\section*{Lead Paragraph}
\textbf{Many physical and biological systems described by nonlinear partial differential equations are difficult to analyze directly. A frequently used approach in reduced order modeling is to track a few key quantities, called “moments,” that summarize the system’s overall behavior at a low-dimensional level. However, accurately modeling how these moments evolve over time usually requires deep mathematical 
closure insights and detailed knowledge of the system. In this work, we introduce a data-driven method based on Neural Ordinary Differential Equations (Neural ODEs) that can automatically learn simplified models from raw simulation or experimental data, even when that data is sparse or unevenly sampled. Our method applies broadly: it can recover governing equations when possible, discover hidden coordinate transformations when needed, and provide reliable approximations when exact models do not exist. This unified Neural ODE framework offers a flexible and powerful alternative to symbolic regression methods like SINDy, with potential implications 
towards model reduction in complex systems in physics, biology, and beyond.}}

\section{Introduction}

There exist numerous nonlinear partial differential equations
in dispersive, as well as in dissipative systems which 
are of broad applicability in a wide range of 
spatio-temporally dependent physical and biological settings.
For instance, on the dispersive side, the Nonlinear Schr{\"o}dinger 
(NLS) model~\citep{Sulem,AblowitzPrinariTrubatch}
has been argued to be of relevance to optical~\citep{Kivshar2003,hasegawa:sio95} and
atomic physics~\citep{becbook2,pethick,siambook}
to plasma~\citep{kono,infeld} as well as fluid 
research~\citep{ablowitz2,infeld} and even to 
biological applications such as the DNA 
denaturation~\cite{PRL_peyrard,Peybi}.
{This work has also been central to the
seminal contributions of S. Aubry, especially
in connection to discrete solitons and
breathers~\cite{MagnusJohansson_1997,Aubry2000,Aubry06}.}
In a similar vein, in dissipative, reaction-diffusion
systems one of the central models has been
the Fisher-KPP (FKPP) equation which was originally conceived
in the context of spatial spread of advantageous alleles,
and has since been employed to model species invasion, population dispersal, and ecological front propagation~\cite{Murray2002,jackxin}. The FKPP model
has also constituted for almost a century now a hallmark 
of reaction-diffusion models with a wide range
of applications to cancer modeling, wound healing,
flame propagation and a diverse further host of
applications up to this day~\cite{Simpson2024,Contri2018}

While such classical PDE models have for decades now been
at the center of initially analytical and subsequently
computational studies, over the past decade or so, a new
suite of data-driven tools and techniques has met with
explosive growth offering a new avenue and a fresh perspective
enabling unprecedented developments. In particular,
the proposal of physics-informed
neural networks (PINNs) by \citep{raissi_physics-informed_2019}
and extensions thereof such as DeepXDE~\citep{lu2021deepxde},
and the essentially concurrent track of sparse identification
of nonlinear systems, so-called SINDy by \citep{Kutz_SINDy}
have arguably constituted ``game-changers'' towards
the analysis of nonlinear ODE and PDE systems.
These have been rapidly followed by 
the sparse optimization of \citep{schaeffer2017learning}, the meta-learning of~\citep{feliu2020meta}, and the neural operators of~\citep{li2021fourier} opening new vistas towards both
the direct problem of solving ODEs and PDEs, as well as
perhaps more importantly towards the inverse problem
of discovery thereof, notably not only from synthetic or
computational, but also potentially from experimental data.
A relevant summary of such model identification techniques 
has been given, e.g., 
in~\citep{karniadakis2021physics}.
An important parallel track that such techniques empower
is toward the identification of conservation laws
inherent in the systems~\citep{teg1,teg2,liu2022machine,PhysRevE.108.L022301} 
and eventually in their potential integrability~\citep{kripy,DEKOSTER2024103273}, including,
e.g., via the so-called Lax pairs~\cite{SILO}.

In a recent work, we have revisited from
a data-driven perspective an intriguing connection
between ODE models and nonlinear PDEs through the
so-called moment methods~\cite{10.3389/fphot.2024.1444993}.
This class of methods~\citep{victor_theory,Victor_NLS_equation}.
allow for the derivation of closed, low-dimensional dynamical
systems. Operating as a physically-inspired reduced-order
model (or manifold of projection of the PDE dynamics), they
provide
insights on the evolution of relevant degrees of freedom
such as the center of mass,
variance, kurtosis etc. of an effective ``distribution''
described by the relevant field theoretic (spatio-temporal) model.
Importantly, such approaches have been successful in the bibliography beyond the confines of dispersive models
(as in the references above), being also applicable to
FKPP models, where they have been considered, e.g., in the
context of brain tumors~\citep{BELMONTEBEITIA20143267}.

In our previous work in this vein~\cite{10.3389/fphot.2024.1444993},
we have leveraged the widely-used 
 SINDy toolbox~\cite{Kutz_SINDy} towards deriving the
 moment closure. SINDy relies on estimating time derivatives, 
 which, in turn, requires the PDE solution to be observed at finely spaced time intervals so that finite differences can accurately approximate the derivatives. Such a requirement limits the method’s applicability in realistic and practical settings, where the underlying PDE model may be unknown and only sparse observations of experimental data are available. Additionally, derivative estimation is well-known to be highly sensitive to noise~\cite{tang2023weakident}, which further undermines the robustness of the method in real-world scenarios where measurement uncertainty is unavoidable.

To overcome these issues, we turn to \textit{Neural Ordinary Differential Equations} (Neural ODEs) by Chen et al.~\cite{chen2018neural} as an alternative, more robust 
approach. Here, the unknown right-hand side of the reduced moment system is represented by a neural network
\begin{equation}
\label{eq:Neural_ODE_intro}
\frac{d\mathbf{x}}{dt} = f(\mathbf{x}, t; \bm{\Theta}),
\end{equation}
where $\bx=\bx(t)$ is the evolution of the selected moments and the parameters $\mathbf{\Theta}$ are learned so that the \textit{integrated} trajectory $\widehat{\bx}(t)$ matches the observed moments rather than their numerically estimated derivatives. This loss formulation eliminates the need for densely sampled, low-noise data and allows us to train on the sparse, irregular time series that  naturally arise when moments are extracted from experiments or large-scale simulations. Once fitted, the Neural ODE yields a continuous-time surrogate that
features the following advantages:
\begin{enumerate}
    \item \textbf{It acts like a conventional ODE}, permitting standard dynamical-systems analysis (conserved quantities, stability, long-time behavior) for the low-dimensional moment description of the PDE;
    \item \textbf{It combines naturally with Riemannian optimization} (e.g., on the Stiefel manifold) to discover linear coordinate transformations that \textit{close} moment systems that are otherwise unclosed in the original coordinates; and
    \item \textbf{It offers a principled reduced-order model} even when no exact closure exists, yielding an accurate, data-driven approximation to the full PDE dynamics.
\end{enumerate}
All of these traits will be evident in our illustrations
that follow.

Since the seminal work of Chen et al. \cite{chen2018neural}, Neural ODEs have become a versatile tool for data-driven discovery of continuous-time dynamics.  Recent applications span a wide spectrum: learning from irregular and partially observed time series \cite{huang2020learning}, modelling quantum dynamics with latent Neural ODEs \cite{choi2022learning}, inferring interactions on dynamical graphs \cite{huang2021coupled, asikis2022neural}, and biological problems ranging from tumour-growth forecasting \cite{laurie2023explainable} to genome-wide regulatory dynamics \cite{hossain2024biologically}. Their reduced-order modeling capabilities have also been demonstrated for combustion chemical kinetics \cite{dikeman2022stiffness}, fluid flows \cite{rojas2021reduced}, and spatio-temporal chaos \cite{linot2022data}.

We push this line of research into the moment-closure setting. Our goal is not only to learn a low-dimensional surrogate, but to (i) exactly recover the governing moment equations whenever a closed-form system exists for the selected moments, and (ii) discover a coordinate transformation that closes the dynamics when such a system does not exist in the original variables.  In doing so, we obtain interpretable, finite-dimensional, closed-form moment descriptions of the underlying infinite-dimensional PDE directly from sparse or irregular observations.
Our presentation is structured as follows: in 
section~\ref{sec:background}, we provide the modeling
context, specific examples and earlier work
motivating our analysis. In section~\ref{sec:method}, 
we discuss the details of our Neural ODE approach.
In section~\ref{sec:experiments}, we expand on our
numerical experiments for the different case examples.
Finally, in section~\ref{sec:conclusion}, we
summarize our conclusions and provide some directions
for future study.

\section{Background and motivation}
\label{sec:background}
This section begins with brief introduction of the two representative PDE systems central to this work. We then introduce the method of moments, which enables reduced-order modeling by transforming PDEs into (potentially closed) systems of ODEs. Finally, we discuss the strengths and, 
importantly, the limitations of existing data-driven moment methods, such as those relying on SINDy~\cite{10.3389/fphot.2024.1444993}, which motivate the approach developed in this paper.

\textbf{Nonlinear Schr\"{o}dinger equation.} The first PDE system we consider is a specific case of the (1+1)-dimensional nonlinear Schr\"{o}dinger (NLS) equation with a harmonic trapping potential, $V(x, t) = \frac{1}{2}x^2$~\cite{siambook,Kivshar2003},
\begin{equation}
\label{eq:nls_harmonic}
    iu_t = -\frac{1}{2}u_{xx} + \frac{1}{2}x^{2}u + g\left(\left|u\right|^{2}, t\right)u,
\end{equation}
where $g\left(|u|^{2},t\right)$ captures the system's nonlinearity. The quadratic confinement term arises in a wide range of applications, from graded-index profiles in nonlinear fiber optics~\cite{Kivshar2003} to the widely used magnetic trapping in dilute atomic Bose–Einstein condensates~\cite{becbook2,siambook}.

\textbf{Fisher-Kolmogorov equation.} The second system we consider is the Fisher–Kolmogorov-Petrovskii-Piskounov (FKPP) reaction–diffusion model
\begin{equation}
\label{eq:fk_equation}
    u_t = D u_{xx}+ \rho u(1-u)
\end{equation}
where $u(x, t)$ denotes the population density normalized by the carrying capacity. The diffusion term $D u_{xx}$ with $D > 0$ represents spatial dispersal, while the logistic reaction term $\rho u(1 - u)$ with intrinsic growth rate $\rho > 0$ describes local reproduction constrained by limited resources. This equation captures the essential interplay between dispersal and growth and serves as a canonical model for invasion fronts, traveling waves, and pattern formation in mathematical biology \cite{Murray2002,shigesada1997biological,volpert2009reaction,Contri2018,Simpson2024,BELMONTEBEITIA20143267}.

In this paper, we will consider initial value problems for Eq.~\eqref{eq:nls_harmonic} and Eq.~\eqref{eq:fk_equation}, assuming localized and sufficiently regular initial data $u_0(x) =u(x, 0)$, so that the relevant moment quantities (introduced below) are well-defined for all $t\ge 0$.

\subsection{The method of moments}
The scope of the \textit{method of moments}~\cite{victor_theory} is
to derive from the PDE a system of ODEs characterizing the evolution of certain spatial integral quantities---referred to as \textit{moments}---of the solution $u(x, t)$. The resulting system of ODEs offers a reduced-order description that captures key features of the original PDE, such as the center of mass motion, the spreading (variance), or shape deformation of the solution. The method has been widely applied in fluid dynamics, kinetic theory, population biology, and nonlinear optics, particularly when the solution is expected to remain localized or approximately 
self-similar~\cite{victor_theory}.

As an example, let us consider the NLS equation~\eqref{eq:nls_harmonic}. Following the formulation in~\cite{victor_theory}, relevant  moments of a solution $u(x, t)$ to Eq.~\eqref{eq:nls_harmonic} are given by:
\begin{align}
\label{eq:def_IVKJ}
\left\{
    \begin{aligned}
        I_{k}(t) &= \int_{\mathbb{R}} x^{k}|u(x,t)|^{2}dx,\\
        V_{k}(t) &= 2^{k-1}i\int_{\mathbb{R}}  x^{k}\left(u(x,t)\frac{\partial \bar{u}(x,t)}{\partial x} - \bar{u}(x,t)\frac{\partial u(x,t)}{\partial x}\right)dx,\\
        K(t) &= \frac{1}{2}\int_{\mathbb{R}}  \left|\frac{\partial u(x,t)}{\partial x}\right|^{2} dx,\\
        J(t) &= \int_{\mathbb{R}}  G\left(\rho(x, t), t\right)dx = \int_{\mathbb{R}}  G\left(\left|u(x, t)\right|^2, t\right)dx
    \end{aligned}\right.
\end{align}
Here, $\bar{u}$ denotes the complex conjugate of $u$, $\rho(x, t) = |u(x, t)|^{2}$ is the mass density, and $G(\rho, t)$ is the antiderivative of $g(\rho, t)$ with respect to $\rho$, i.e., $\frac{\partial G}{\partial \rho}(\rho, t) = g(\rho, t)$. These moments have meaningful physical interpretations: for instance, $I_1$ represents the center of mass of the density $\rho = |u|^2$, and $V_1$ is the corresponding moment associated with the momentum density. The quantities $I_k$ and $V_k$, $k\ge 2$, are higher-order moments (e.g., variance, skewness, kurtosis, etc.e) with respect to the corresponding distributions. $K$ could be interpreted as the kinetic energy, while $J$ accounts for the contribution from the nonlinear part of the energy.

Under specific choices of the nonlinearity $g(\rho, t)$, it is possible to \textit{analytically} derive a closed system of ODEs governing the evolution of (a subset of) the moments~\cite{victor_theory,Victor_NLS_equation}. Below, we present two illustrative examples for the NLS equation~\eqref{eq:nls_harmonic}.

\begin{example}
\label{ex:ex_1}
In the linear regime of Eq.~\eqref{eq:nls_harmonic}, where $g(\rho, t) \equiv 0$, the moments ${I_2, V_1, K}$ evolve according to a closed system of ODEs:
    \begin{align}
    \label{eq:second_ME}
    \left\{
    \begin{aligned}
        \frac{dI_2}{dt} &= V_1,\\
        \frac{dV_1}{dt} &= 4K - 2I_2,\\
        \frac{dK}{dt} &= -\frac{1}{2}V_1.
    \end{aligned}\right.
    \end{align}
\end{example}

\begin{example}
\label{ex:ex_2}
Consider the case where the nonlinearity is time-independent, i.e., $g(\rho, t) = g(\rho) = g_0 \rho^2$, where $g_0 \in \mathbb{R}$ is a constant. If we choose the state variables to be the moments $I_2$, $V_1$, $K$, and $J$, their evolution does not form a closed system of ODEs.

However, a straightforward calculation shows that the system becomes closed under the coordinate transformation $E = K + J$:
    \begin{align}
    \label{eq:third_ME}
    \left\{
    \begin{aligned}
        \frac{dI_2}{dt} &= V_1,\\
        \frac{dV_1}{dt} &= 4E - 2I_2,\\
        \frac{dE}{dt} &= -\frac{1}{2}V_1,
    \end{aligned}\right.
    \end{align}
\end{example}

In addition to the above analytic moment closures available for the NLS equation~\eqref{eq:nls_harmonic}, one can also derive \textit{approximate} closures for the FK equation~\eqref{eq:fk_equation} using the so-called ``effective particle methods''. Consider a traveling-wave ansatz of the form
\begin{align}
\label{eq:travel_ansatz}
    u(x, t) = \frac{A(t)}{[1+e^{(x-X(t))/w(t)}]^2},
\end{align}
in line with the analysis of~\cite{BELMONTEBEITIA20143267},
where $A(t)$ denotes the wave amplitude, and $X(t)$ and $w(t)$  represent the front position and width, respectively. Define the following integral quantities:
\begin{align}
    \label{eq:FK_eq}
    \left\{
    \begin{aligned}
        \mathcal{I}_1 &= \int_{-\infty}^{\infty}u_x\,dx,\\
        \mathcal{I}_2 &= \frac{\int_{-\infty}^{\infty}xu_x\,dx}{\mathcal{I}_1(t)},\\
        \mathcal{I}_3 &= \frac{\int_{-\infty}^{\infty}(x - \mathcal{I}_2(t))^2 u_x\,dx}{\mathcal{I}_1(t)},
    \end{aligned}\right.
\end{align}
which correspond, respectively, to the number of particles, center of mass and width of the gradient density $u$. Straightforward calculations~\cite{BELMONTEBEITIA20143267} yield the following evolution equations.

\begin{example}
\label{ex:ex_3}
If $u(x, t)$ of the form~\eqref{eq:travel_ansatz} is a solution to the FK equation~\eqref{eq:fk_equation}, then the moments $\mathcal{I}_i$, $i=1,2, 3$, defined in Eq.~\eqref{eq:FK_eq}, evolve according to the following closed system of ODEs:
\begin{align}
    \label{eq:third_ME_2}
    \left\{
    \begin{aligned}
        \frac{d\mathcal{I}_1}{dt} &= \rho \mathcal{I}_1(1+\mathcal{I}_1),\\
        \frac{d\mathcal{I}_2}{dt} &= -\frac{5}{6}\rho\sqrt{\frac{3}{\pi^{2}-3}}\mathcal{I}_1\sqrt{\mathcal{I}_3},\\
        \frac{d\mathcal{I}_3}{dt} &= 2D + \frac{\rho}{\pi^{2}-3}\mathcal{I}_1\mathcal{I}_3.
    \end{aligned}\right.
\end{align}
\end{example}

\begin{remark}
\label{remark:approx_closure}
    Unlike Examples~\ref{ex:ex_1} and~\ref{ex:ex_2}, where the moment systems~\eqref{eq:second_ME} and~\eqref{eq:third_ME} hold for all solutions that are sufficiently regular to ensure the existence of moments for all $t \ge 0$, the moment system in Eq.~\eqref{eq:third_ME_2} is valid only for solutions $u(x,t)$ of the FKPP equation that exactly take the form of the ansatz~\eqref{eq:travel_ansatz}. Since $u(x,t)$ only approximately satisfies this form, Eq.~\eqref{eq:third_ME_2} provides only an approximate description of the moment dynamics.
\end{remark}

\subsection{Data-driven discovery of moment equations via SINDy}
\label{sec:sindy}

Although there are scenarios where the method of moments leads to analytically closed systems, obtaining such closures typically relies on deep knowledge of the governing PDEs and careful symbolic derivation. To bypass these constraints, our previous work~\cite{10.3389/fphot.2024.1444993} proposed a data-driven approach for discovering exact or approximate moment closure systems directly from numerical simulations, without requiring explicit analytical treatment.

The key methodology in that work is the Sparse Identification of Nonlinear Dynamics (SINDy) algorithm~\cite{Kutz_SINDy}. Given a sequence of PDE solutions $\left\{u(x, t_1), \cdots, u(x, t_N)\right\}$,  we first compute a corresponding time series of moment values by integrating in space using formulas such as Eq.~\eqref{eq:def_IVKJ} or Eq.~\eqref{eq:FK_eq}. For example, in Example~\ref{ex:ex_1}, this produces the \textit{state matrix}
\begin{align}
    \bX =
    \begin{bmatrix}
        | & | & | \\
        \mathbf{I}_2 & \mathbf{V}_1 & \mathbf{K}\\
        | & | & |
    \end{bmatrix}
    =
    \begin{bmatrix}
        I_2(t_1) & V_1(t_1) & K(t_1)\\
        I_2(t_2) & V_1(t_2) & K(t_2)\\
        \vdots & \vdots & \vdots\\
        I_2(t_N) & V_1(t_N) & K(t_N)
    \end{bmatrix}\in\R^{N\times 3}
\end{align}
Next, we approximate the time derivatives of these moments using finite differences, resulting in the \textit{derivative matrix} $\dot{\bX}\in\R^{N\times 3}$:
\begin{align}
\label{eq:derivative_matrix}
    \frac{d}{dt}\bX = 
    \dot{\bX} =
    \begin{bmatrix}
        | & | & | \\
        \dot{\mathbf{I}}_2 & \dot{\mathbf{V}}_1 & \dot{\mathbf{K}}\\
        | & | & |
    \end{bmatrix}
    =
    \begin{bmatrix}
        \dot{I}_2(t_1) & \dot{V}_1(t_1) & \dot{K}(t_1)\\
        \dot{I}_2(t_2) & \dot{V}_1(t_2) & \dot{K}(t_2)\\
        \vdots & \vdots & \vdots\\
        \dot{I}_2(t_N) & \dot{V}_1(t_N) & \dot{K}(t_N)
    \end{bmatrix}\in\R^{N\times 3}
\end{align}
To model these time derivatives, we construct a dictionary of  candidate functions built from the observed moment values. For example, a quadratic dictionary takes the form
\begin{align}
\label{eq:quadratic_dictionary}
\mathcal{D}_{\deg =1, 2}(\mathbf{X}) =
\Big[ \underbrace{\mathbf{I}_2,\ \mathbf{V}_1,\ \mathbf{K}}_{\text{degree 1}},\
\underbrace{\mathbf{I}_2^2,\ \mathbf{V}_1^2,\ \mathbf{K}^2,\ \mathbf{I}_2 \mathbf{V}_1,\ \mathbf{I}_2 \mathbf{K},\ \mathbf{V}_1 \mathbf{K}}_{\text{degree 2}} \Big]
\in \mathbb{R}^{N\times 9}.
\end{align}
SINDy then solves a sparse regression problem, fitting each column of $\dot{\bX}$ as a sparse linear combination of the terms in $\mathcal{D}_{\deg =1, 2}(\mathbf{X})$. This yields interpretable models for the evolution of the moments.

While \cite{10.3389/fphot.2024.1444993} demonstrated successful recovery of moment dynamics for NLS equations---including cases with exact closures (Example~\ref{ex:ex_1}) and those requiring coordinate transformations (Example~\ref{ex:ex_2})---the approach has intrinsic limitations. In particular,
\begin{itemize}
    \item The SINDy method relies on estimating time derivatives, as described in Eq.~\eqref{eq:derivative_matrix}. This requires the PDE solution to be observed at finely spaced time intervals so that finite differences can accurately approximate the derivatives. Such a requirement limits the method’s applicability in realistic and practical settings, where the underlying PDE model may be unknown and only sparse observations of experimental data are available. Additionally, derivative estimation is highly sensitive to noise~\cite{tang2023weakident}, which further undermines the robustness of the method in real-world scenarios where measurement uncertainty is unavoidable.
    \item Because the SINDy loss function is imposed on the local-in-time mismatch between estimated derivatives and sparse combinations of dictionary terms---rather than aiming to match the long-time evolution of the moments---the recovered models can exhibit qualitatively incorrect behavior. For example, they may produce solutions that blow up in finite time, even when the true moment dynamics remain globally well-defined; see Section 4.2.2 of \cite{10.3389/fphot.2024.1444993}.
\end{itemize}

\section{Proposed Method}
\label{sec:method}
Motivated by the limitations of~\cite{10.3389/fphot.2024.1444993}, we propose a data-driven approach for discovering moment dynamics using Neural Ordinary Differential Equations (Neural ODEs)~\cite{chen2018neural}. We first show how Neural ODEs can overcome key drawbacks of the SINDy method by minimizing discrepancies in the moment values themselves, rather than relying on time derivative estimates. We then demonstrate how the Neural ODE framework can be extended to learn coordinate transformations that close the  moment dynamics, even when closure does not hold in the original coordinates.

\subsection{Neural ODE trained on sparsely sampled observational data}
\label{sec:neural_ode_method}

Introduced by \cite{chen2018neural}, Neural ODEs offer a powerful framework for modeling systems with inherently continuous dynamics, making them particularly well-suited for applications in physics and scientific machine learning.  In this framework, the evolution of a state variable $\mathbf{x}(t) \in \mathbb{R}^n$ is modeled by parameterizing the right-hand side of an ODE with a neural network:
\begin{equation}
\label{eq:Neural_ODE}
\frac{d\mathbf{x}}{dt} = f(\mathbf{x}, t; \bm{\Theta}),
\end{equation}
where $\mathbf{x}(t) = (x_1(t), \cdots, x_n(t))^\top$ denotes the system state, and $f(\cdot, \cdot; \bm{\Theta}) : \mathbb{R}^n \times \mathbb{R} \to \mathbb{R}^n$ is a neural network parameterized by $\theta$ that defines the system dynamics.

Given an initial condition $\mathbf{x}(t_0)$, the solution at a later time $t_1$ is obtained by solving the initial value problem:
\begin{equation}
\label{eq:ODE_integration}
\widehat{\mathbf{x}}(t_1; \bm{\Theta}) = \mathbf{x}(t_0) + \int_{t_0}^{t_1} f(\mathbf{x}(t), t; \bm{\Theta})\, dt \eqqcolon F(\bx(t_0), t_0, t_1; \mathbf{\Theta}),
\end{equation}
where the integral is approximated using a numerical ODE solver. This formulation allows the model to learn and approximate the underlying dynamical system from data.

When the system is observed at discrete time points $t_1, \cdots, t_M$---not necessarily closely spaced---the Neural ODE can be trained by minimizing the discrepancy between the predicted and observed state values using the loss function:
\begin{equation}
\label{eq:neural_ode_loss_background}
\mathcal{L}_{\text{Neural ODE}}(\bm{\Theta}) = \sum_{i=1}^{M} \left| \mathbf{x}(t_i) - \hat{\mathbf{x}}(t_i; {\bm{\Theta}}) \right|^2,
\end{equation}
This avoids the need to estimate time derivatives, which can be noisy and unreliable in sparse or experimental settings.

Instead of integrating from a single initial time $t_0$, it is often more effective to train on multiple short trajectory segments using mini-batches. This improves data efficiency and generalization. We define two parameters: the \textit{batch size $L$} (number of sampled initial times) and the \textit{batch length $T$} (number of time steps per segment). Each batch consists of $L$ short trajectories $\{\mathbf{x}(t_{i_l}), \cdots, \mathbf{x}(t_{i_l + T - 1})\}_{l=1}^L$, each starting from a different randomly selected $t_{i_l}$ with initial state $\mathbf{x}(t_{i_l})$.  These are then integrated forward using the Neural ODE. The corresponding mini-batch loss is:
\begin{equation}
\label{eq:neural_ode_loss_batch}
\mathcal{L}_{\text{Neural ODE}}(\bm{\Theta}) = \frac{1}{LT} \sum_{l=1}^{L} \sum_{j=1}^{T} \left\| \mathbf{x}(t_{i_l + j}) - \widehat{\mathbf{x}}(t_{i_l + j}; \bm{\Theta}) \right\|^2,
\end{equation}
where $\widehat{\mathbf{x}}$ denotes the predicted trajectory obtained by integrating from $\mathbf{x}(t_{i_l})$ using Eq.~\eqref{eq:ODE_integration}.

In this work, we adopt the Neural ODE framework to learn the dynamics of the moment system. To illustrate the approach, we consider Example~\ref{ex:ex_1}, where the state variable is defined as
\[
\mathbf{x} = \begin{bmatrix} I_2, V_1,K \end{bmatrix}^\top \in \mathbb{R}^{3}
\]
To model the right-hand side of the moment system~\eqref{eq:second_ME}, we first apply a nonlinear feature transformation $\mathcal{D} : \mathbb{R}^{3} \to \mathbb{R}^{d}$ that lifts the input into a higher-dimensional space, analogous to selecting a dictionary in SINDy. For instance, using a quadratic feature map, as in Section~\ref{sec:sindy}, the transformation $\mathcal{D}_{\deg =1, 2}(\mathbf{X})$ includes both first- and second-degree interactions:
\begin{align}
\label{eq:quad_feature_map}
\mathcal{D}_{\deg =1, 2}(\mathbf{x}) =
\Big[ \underbrace{I_2,\ V_1,\ K}_{\text{degree 1}},\
\underbrace{I_2^2,\ V_1^2,\ K^2,\ I_2 V_1,\ I_2 K,\ V_1 K}_{\text{degree 2}} \Big]^\top
\in \mathbb{R}^{9}.
\end{align}
The time evolution of $\mathbf{x}$ is then modeled by a linear projection in this feature space:
\begin{equation}
\label{eq:linear_proj}
\frac{d\mathbf{x}}{dt} = f(\mathbf{x}(t), t; \boldsymbol{\Theta}) 
\approx \boldsymbol{\Theta} \cdot \mathcal{D}_{\deg =1, 2}(\mathbf{x})
= 
\begin{bmatrix}
\text{--- } \bm{\theta}_1^\top \text{ ---} \\
\text{--- } \bm{\theta}_2^\top \text{ ---} \\
\text{--- } \bm{\theta}_3^\top \text{ ---}
\end{bmatrix}
\cdot
\begin{bmatrix}
I_2 \\
V_1 \\
\vdots \\
V_1K
\end{bmatrix}
\in \mathbb{R}^{3 \times n},
\end{equation}
where $\mathbf{\Theta}\in \mathbb{R}^{3 \times 9}$ is a learnable weight matrix. This linear projection~\eqref{eq:linear_proj} corresponds to a single-layer neural network without bias operating on the nonlinear feature $\mathcal{D}(\mathbf{x})$.

Note that the feature transformation $\mathcal{D}$ is not restricted to including only quadratic interaction terms. It can be customized to include the constant term (i.e., zeroth-order term), higher-order polynomial terms such as cubic or quartic interactions, or any other nonlinear combinations depending on the application. Importantly, any change in the dimensionality of $\mathcal{D}(\mathbf{X})$ requires a corresponding adjustment to the dimensions of $\mathbf{\Theta}$ to ensure compatibility.  While enriching the feature space increases expressivity, it also introduces additional complexity and may reduce interpretability.

\textbf{Sparse model discovery via iterative thresholding.}
A key assumption in our framework is that the underlying dynamics \( f(\mathbf{x}) \) admit a \emph{simple}, sparse representation in terms of a predefined feature library \( \mathcal{D}(\mathbf{x}) \). For instance, the right-hand sides of Equations~\eqref{eq:second_ME} and~\eqref{eq:third_ME} can each be expressed as sparse linear combinations of such dictionary terms. Under this assumption, the linear model in Eq.~\eqref{eq:linear_proj} involves a sparse coefficient matrix \( \mathbf{\Theta} \in \mathbb{R}^{3 \times 9} \), where each row \( \bm{\theta}_j \) selects a minimal subset of dictionary terms relevant to the corresponding component \( f_j(\mathbf{x}) \).

To identify such sparse structure, we adopt an iterative thresholding procedure during training. After the initial convergence of the Neural ODE, we apply a hard threshold to $\mathbf{\Theta}$:  entries with magnitude below a user-defined cutoff are set to zero. Gradients corresponding to these pruned entries are then frozen, ensuring that subsequent updates affect only the remaining active terms. To facilitate adaptation to this sparsification, we temporarily increase the learning rate, followed by annealing for fine-tuning. This procedure results in a compact and interpretable model that captures the essential dynamics with minimal complexity.

\begin{figure}[t]
\centering
\includegraphics[width=0.8\linewidth]{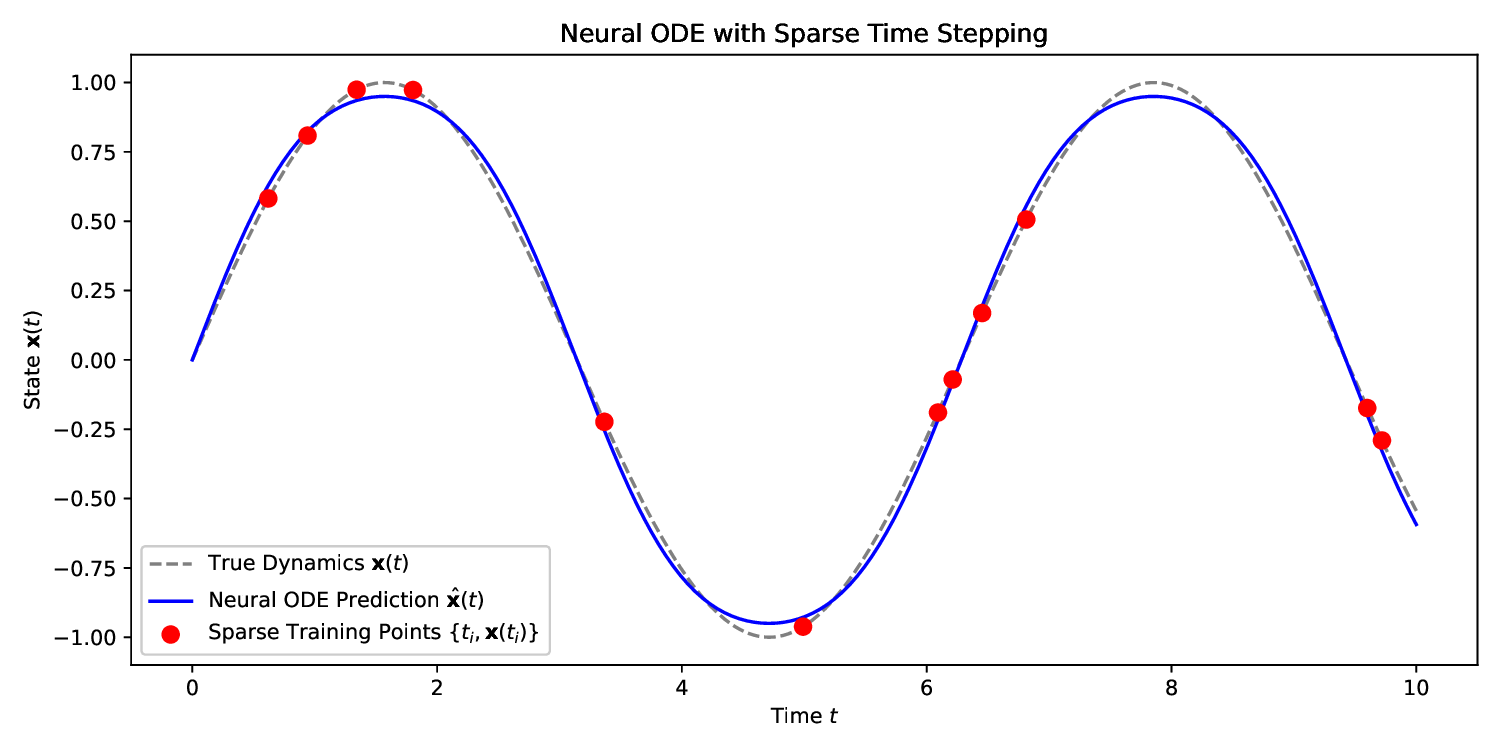}
\caption{Neural ODE trained with sparsely sampled observational data.}
\label{fig:neural_ode_sparse}
\end{figure}

\textbf{Sparsely sampled observational data}. To address the limitations of the SINDy-based approach discussed at the end of Section~\ref{sec:sindy}, we consistently train our model using sparsely sampled time series, rather than relying on densely and uniformly sampled data required for accurate numerical time-differentiation. This setup is illustrated in Figure~\ref{fig:neural_ode_sparse}, where only a small number of observations (red points) are used for training, drawn from the full trajectory of the PDE solution (gray dashed curve).

Notably, our comparisons include the method in~\cite{10.3389/fphot.2024.1444993}, which depends on SINDy applied to fully sampled data---a necessary condition for its performance. We will show in Section~\ref{sec:experiments} that our approach remains robust and outperforms the SINDy-based model even under highly sparse data regimes where SINDy fails entirely.

\subsection{Learning coordinate transformations for moment system closure} \label{sec:stiefel_neural_ode}

In certain systems, the evolution equations for the observed quantities are not closed but may become closed under unknown coordinate transformations. As an illustrative case, we focus on Example~\ref{ex:ex_2}, where the dynamics of the initially selected moments $(I_2, V_1, K, J)$ do not form a closed system. The goal is to identify---through data-driven methods---a coordinate transformation, such as $E = K+J$, that renders the system closed.

To this end, let $\bX = [\mathbf{I}_2, \mathbf{V}_1, \mathbf{K}, \mathbf{J}]\in\R^{N\times 4}$ denote the moment time series observed at (potentially irregular and sparse) time points $t_1, \cdots, t_N$. To identify a linear coordinate transformation $\bx=(I_2, V_1, K, J)^\top\mapsto\mathbf{y}=\mathbf{A}^\top\bx$, where $\mathbf{A}\in\R^{4\times 3}$ that yields a closed system, we formulate the following optimization problem:
\begin{align}
\label{eq:optimization_inner_outer}
    & \min_{\bA\in \R^{4\times 3}}\min_{\mathbf{\Theta}\in\R^{3\times |\mathcal{D}|}} \sum_{i=1}^{N-1}\left\|F(
    (\bX\bA)[i, :], t_i, t_{i+1}; \mathbf{\Theta})- (\bX\bA)[i+1,:]  \right\|^2 \\
    & \text{ s.t.~}   \mathbf{A}^\top \mathbf{A} = \mathbf{I}_{3\times 3}  
\end{align}
where $\mathbf{\Theta}\in\R^{3\times |\mathcal{D}|}$ is a \textit{sparse} coefficient matrix (with respect to a prescribed dictionary $\mathcal{D}$) parameterizing the Neural ODE dynamics in the reduced coordinates. The function $F(\bx_0, t_i, t_{i+1}; \mathbf{\Theta})$ denotes the Neural ODE solver that integrates the system forward from $t_i$ to $t_{i+1}$ with initial condition $\bx_0$; see Eq.~\eqref{eq:ODE_integration} for its definition. The matrix $\bX\bA \in \mathbb{R}^{N \times 3}$ is the state matrix of the transformed coordinates, and $(\bX\bA)[i+1, :] \in\mathbb{R}^{1 \times 3}$ denotes the $(i+1)$-th row.

\begin{remark}
    The idea behind Eq.~\eqref{eq:optimization_inner_outer} is closely related to the approach in~\cite{10.3389/fphot.2024.1444993}. Specifically, the goal is to find a coordinate transformation $\bx\mapsto\bA^\top \bx$ such that the dynamics in the transformed coordinates can be described by a closed, parsimonious ODE. The constraint $\bA^\top\bA = \mathbf{I}_{3\times 3}$ eliminates the trivial solution $\bA = \mathbf{0}$. The set of matrices satisfying this constraint forms the \textit{Stiefel manifold}~\cite{edelman1998geometry,absil2008optimization}. 
\end{remark}

\begin{algorithm}[hbt!]
\caption{Data-driven discovery of coordinate transformation}
\label{alg:inner_outer_optimization}
\KwInput{\texttt{maxIter}: Total number of iterations.\\
\hspace{4.5em}\texttt{$lr_{initial}$}: Initialized learning rate;\quad \texttt{$lr_{scheduled}$}: Scheduled learning rate.\\
\hspace{4.5em}\texttt{$Iter_{scheduled}$}: Number of initial iterations during which $lr$ remains fixed.\\
\hspace{4.5em}Initializations: $\mathbf{A}^{(0)}\in\mathbb{R}^{4\times 3}$, $\mathbf{\Theta}^{(0)}\in\mathbb{R}^{3\times |\mathcal{D}|}$.}
\KwOutput{$\mathbf{A}_{\text{out}}$, $\mathbf{\Theta}_{\text{out}}$}
 
\For{$k = 1$ \KwTo $\texttt{maxIter}$}{
    \If{$k > \texttt{Iter}_{\texttt{scheduled}}$}{
        Update learning rate: $\texttt{lr}_{\texttt{scheduled}} \gets \texttt{lr}_{\texttt{initial}}$\;
    }
    
    \textbf{Inner loop:} optimize $\mathbf{\Theta}^{(k)}$ with fixed $\mathbf{A}^{(k-1)}$ using mini-batch Neural ODE training: \\
    \Indp For each mini-batch (in time) $\{ t_{i_j}\}_{j=1}^M$, solve
    \[
    \mathbf{\Theta}^{(k)} \leftarrow \arg\min_{\mathbf{\Theta}}  \sum_{j=1}^{M-1}\left\|F(
    (\bX\bA)[i_j, :], t_{i_j}, t_{i_{j+1}}; \mathbf{\Theta})- (\bX\bA)[i_{j+1},:]  \right\|^2
    \]
    \Indm
    
    \textbf{Outer loop:} optimize $\mathbf{A}^{(k)}$ with fixed $\mathbf{\Theta}^{(k)}$ under the Stiefel constraint:
    \[
    \mathbf{A}^{(k)} \leftarrow \arg\min_{\mathbf{A}^\top \mathbf{A} = \mathbf{I}}  \sum_{i=1}^{N-1}\left\|F(
    (\bX\bA)[i, :], t_i, t_{i+1}; \mathbf{\Theta})- (\bX\bA)[i+1,:]  \right\|^2
    \]
}
\Return{$\mathbf{A}_{\text{out}} \leftarrow \mathbf{A}^{(\texttt{maxIter})}$, \quad $\mathbf{\Theta}_{\text{out}} \leftarrow \mathbf{\Theta}^{(\texttt{maxIter})}$}
\end{algorithm}

To solve Eq.~\eqref{eq:optimization_inner_outer}, we adopt an alternating optimization strategy:
\begin{itemize}
    \item When $\mathbf{A}$ is fixed, optimizing with respect to $\mathbf{\Theta}$ reduces to a standard Neural ODE learning problem.
    \item When $\mathbf{\Theta}$ is fixed,  the optimization over $\bA$ becomes a constrained problem on the Stiefel manifold, which can be efficiently solved using established Riemannian optimization techniques~\citep{10.1007/978-3-030-33749-0_20, doi:10.1080/10556788.2020.1852236, Liu2021API}.
\end{itemize}

This alternating scheme iteratively updates $\bA$ and $\mathbf{\Theta}$ to jointly discover a low-dimensional coordinate system in which the moment dynamics form a sparse and closed ODE system. See Algorithm~\ref{alg:inner_outer_optimization} for the pseudocode implementing this alternating scheme.

\section{Numerical experiments}
\label{sec:experiments}

In this section, we present numerical results for the data-driven identification and closure of moment systems corresponding to Examples~\ref{ex:ex_1}–\ref{ex:ex_3}, using the methodologies introduced in Section~\ref{sec:method}. The general training configuration used in our experiments is summarized in Table~\ref{tab:Neural_ODE_setting} and remains consistent across all examples unless stated otherwise.

For the NLS equation in Examples~\ref{ex:ex_1} and~\ref{ex:ex_2}, we compare our results against both ground-truth analytical solutions and the numerical results reported in~\cite{10.3389/fphot.2024.1444993}. In particular, we assess the performance of the Neural ODE model and the SINDy-based approach from~\cite{10.3389/fphot.2024.1444993} in identifying governing equations for the moment dynamics and discovering coordinate transformations that yield closure. This comparison is carried out under both densely sampled and, more crucially, sparsely sampled time series data, where the SINDy method fails entirely.

Training data are generated by numerically solving the NLS equation~\eqref{eq:nls_harmonic} with periodic boundary conditions and a single initial condition, using the exponential time-differencing fourth-order Runge–Kutta method (ETDRK4)~\cite{4thordeRK4}. The associated moment time series are computed via spatial integration, as defined in Eq.~\eqref{eq:def_IVKJ}, with spatial derivatives evaluated using pseudo-spectral Fourier methods. This same procedure may also be applied to experimental measurements in place of simulated data. Unless otherwise noted, the NLS simulations consist of $N$ = 16{,}000 uniformly spaced time points with temporal resolution $\Delta t = 0.0025$. When indicated, we also evaluate our Neural ODE framework using sparsely and irregularly subsampled dataset of $M$ = 4{,}000 time points.

For the FKPP equation in Example~\ref{ex:ex_3}, we aim to identify an approximate governing equation for the selected moments $\mathcal{I}_1, \mathcal{I}_2, \mathcal{I}_3$ using the Neural ODE framework and compare it with the analytically derived approximate moment system. See Remark~\ref{remark:approx_closure} for a discussion on why an exact analytical closure is not available. The data are generated by simulating the FKPP equation~\eqref{eq:fk_equation} with initial conditions resembling the ansatz~\eqref{eq:travel_ansatz}, from which we extract moment time series at $N$ = 300 uniformly spaced time points with $\Delta t = 0.1$. When indicated, we also evaluate our Neural ODE framework using sparsely and irregularly subsampled dataset of $M$ = 100 time points from the first 200 entries.

Note that during the training phase of the Neural ODE, each moment is individually normalized to the range $[-1, 1]$. The model is trained on this normalized data, resulting in a learned dynamical system defined in the normalized coordinate space. To interpret the model outputs in the original physical variables, an inverse transformation is applied using the recorded scaling parameters.

\begin{table}[t]
\begin{center}
\begin{tabular}{lll}  \toprule
Number of Layers & 1  \\ \hline
Batch Size $L$ & 200  \\ \hline
Batch Time $T$ & 10  \\ \hline
Activation Function & None (Linear)  \\ \hline
Optimizer & Adam  \\ \hline
Adam $\beta_1$ & 0.9  \\ \hline
Adam $\beta_2$ & 0.999  \\ \hline
Initial Learning Rate & 1e-2  \\ \hline
Scheduled Learning Rate for Fine-Tuning & 1e-4  \\ \hline
Initial Training Steps & 2000  \\ \hline
Fine-Tuning Steps & 1000   \\ \bottomrule
\end{tabular}
\end{center}
\caption{Hyperparameters used for training the Neural ODE model. No nonlinear activation is applied, as nonlinearity is introduced via the predefined feature map $\bx \mapsto \mathcal{D}(\bx)$ (see Section~\ref{sec:neural_ode_method}). Refer to Eq.~\eqref{eq:neural_ode_loss_batch} for the definitions of batch size ($L$) and batch length ($T$).}
\label{tab:Neural_ODE_setting}
\end{table}

\subsection{Examples with analytically closed moment systems}
\label{sec: neural_ode_closed_system}
We begin with Example~\eqref{ex:ex_1}, previously studied in~\cite{10.3389/fphot.2024.1444993}, where an exact linear closure exists for the selected moments $\bfx = [I_2, V_1, K]^\top$. The moment time series matrix $\bX = [\mathbf{I}_2, \mathbf{V}_1, \mathbf{K}] \in \R^{N \times 3}$ is constructed by numerically solving the PDE~\eqref{eq:nls_harmonic} with the initial condition
\begin{align}
    u^{(1)}(x,0) &= 1.88\exp\left(-\frac{1}{2}\left(x-5\right)^{2}\right).
    \label{ex:ex_1_IC_1}
\end{align}
Since a linear closure is already known, we first apply the Neural ODE model to the moment time series using a degree-1 feature map, $\mathcal{D}_{\deg = 1}(\mathbf{\bx}) = \begin{bmatrix} I_2, V_1,K \end{bmatrix}^\top$. After training with iterative thresholding, we transform the learned dynamics back to the original coordinate system. The resulting coefficient matrix for the degree-1 feature library is:
\[
\boldsymbol{\Theta}_{\deg = 1} = 
\begin{bmatrix}
0.000 &  1.000 &  0.000 \\
-2.000 &  0.000 &  4.000  \\
0.000 & -0.500 &  0.000
\end{bmatrix}
\]
This corresponds to the following system of equations:
\begin{align}
    \left\{
    \begin{aligned}
    \label{eq:ex1_result_degree1}
        \frac{dI_2}{dt} &= 1.000V_1,\\
        \frac{dV_1}{dt} &= - 2.000I_2 + 4.000K ,\\
        \frac{dK}{dt} &= -0.500V_1.
    \end{aligned}\right.
    \end{align}
All coefficients are rounded to three decimal places and show excellent agreement with the ground-truth system in Example~\eqref{ex:ex_1}.

We then expand the feature map to include quadratic terms, leading to:
\begin{align*}
    \mathcal{D}_{\deg =1, 2}(\mathbf{x}) = \begin{bmatrix}
I_2,\ V_1,\ K,\ I_2^2,\ V_1^2,\ K^2,\ I_2 V_1,\ I_2 K,\ V_1 K
\end{bmatrix}^\top.
\end{align*}
By applying the same training procedure, we obtain the following learned coefficient matrix:
\[
\boldsymbol{\Theta}_{\deg = 1, 2} = 
\begin{bmatrix}
0.000 &  1.000 &  0.000 &  0.000 &  0.000&  0.000&  0.000&  0.000&  0.000 \\
-2.000 &  0.000 &  4.000 &  0.000 & 0.000&  0.000&  0.000&  0.000&  0.000\\
0.000 & -0.500 &  0.000 &  0.000 & 0.000&  0.000&  0.000&  0.000&  0.000
\end{bmatrix}
\]
This representation again recovers the same system of equations previously obtained in Eq.~\eqref{eq:ex1_result_degree1}. Notably, despite the expanded feature library, all quadratic terms are eliminated through thresholding, demonstrating the ability of the Neural ODE framework to recover parsimonious and accurate dynamics.
\begin{remark}
In contrast to~\cite{10.3389/fphot.2024.1444993}, where SINDy with a quadratic feature library fails to recover the correct dynamics from a single initial condition, the Neural ODE approach proposed herein accurately identifies the underlying structure from the same limited data. This advantage stems from the difference in loss formulation: SINDy minimizes local-in-time discrepancies between estimated derivatives and sparse dictionary terms, which can lead to unstable models---including
ones featuring finite-time blow-up---even when the true moment dynamics are globally well-defined. In contrast, the Neural ODE framework fits the full trajectory, resulting in stable and physically consistent models.
\end{remark}

To further assess the robustness of our approach in practical settings---where data are observed at sparse and irregular time points rather than simulated from a known PDE---we randomly subsample $M=4{,}000$ time points from the original $N=16{,}000$ steps and retrain the model using the same quadratic feature map $\mathcal{D}_{\deg =1, 2}(\mathbf{x})$. The learned dynamics from this reduced dataset remain nearly identical to the system in Eq.~\eqref{eq:ex1_result_degree1}.

To quantitatively evaluate the relevant performance, we compute the relative mean absolute error (RMAE) between the ground-truth moment trajectories and the predicted moments $(\hat{I}_2, \hat{V}_1, \hat{E})$, obtained by integrating the learned ODE system from the same initial conditions. Errors are reported both in the normalized coordinate space and after rescaling to the original physical units. A summary of the reconstruction results is provided in Table~\ref{tab:relative_errors}.

\begin{table}[t]
\centering
\begin{tabular}{lccccc}
\toprule
\textbf{Quantity} & $I_2$ & $V_1$ & $K$ & Overall (Norm.) & Overall (Orig.) \\
\midrule
\textbf{RMAE} & 0.0077 & 0.0194 & 0.0077 & 0.0116 & 0.0090 \\
\bottomrule
\end{tabular}
\caption{
RMAE for each reconstructed moment variable ($I_2, V_1$, $K$) and the overall error (aggregated over all three moments), computed using a Neural ODE model trained on sparse time-series data ($M=4{,}000$ out of $N = 16{,}000$ points) with the quadratic feature map $\mathcal{D}_{\deg =1, 2}(\mathbf{x})$. ``Norm.'' denotes errors in the normalized coordinate space; ``Orig.'' denotes errors after rescaling to the original physical units.}
\label{tab:relative_errors}
\end{table}

The results demonstrate that the Neural ODE framework remains highly accurate even under sparse sampling---a setting in which the SINDy-based method~\cite{10.3389/fphot.2024.1444993} fails entirely due to its reliance on accurately estimating time derivatives, which is not feasible with sparsely sampled data.

\begin{figure}[t]
    \centering
    \includegraphics[width=1\linewidth]{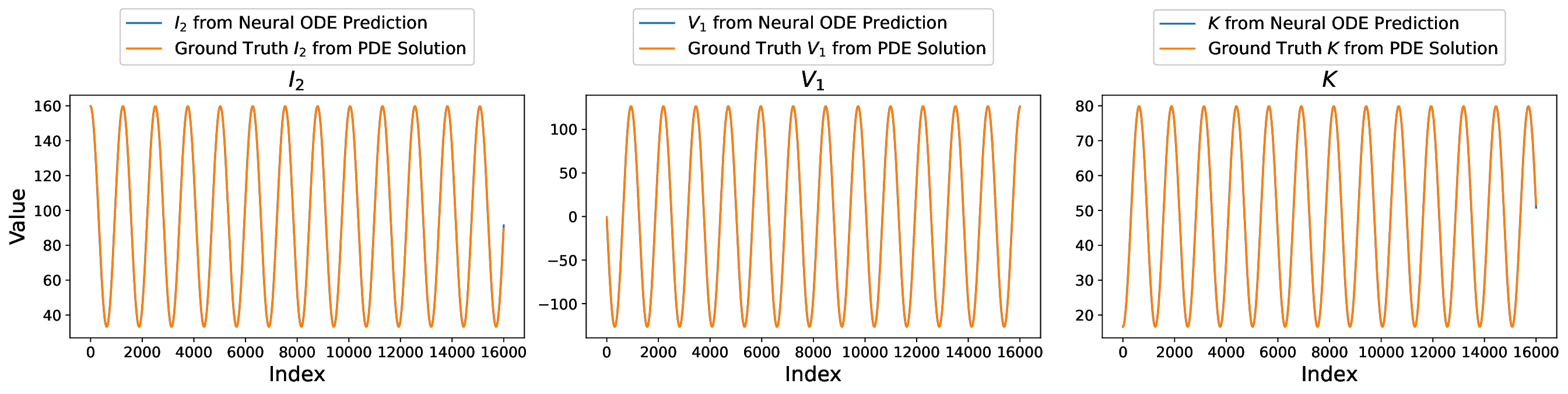}
    \caption{Comparison of ground-truth moment trajectories $[I_2, V_1, K]$ (from PDE simulations) with those reconstructed by Neural ODEs trained on sparse time-series data ($M=4{,}000$ out of $N = 16{,}000$ points), using a quadratic feature library $\mathcal{D}_{\deg=1, 2}(\bx)$. Results from the SINDy-based method~\cite{10.3389/fphot.2024.1444993} are omitted due to its failure under sparse sampling, which stems from its reliance on time-derivative estimation.}
    \label{fig: IVE_sparse_data}
\end{figure}

As a visual illustration, the three panels of Figure~\ref{fig: IVE_sparse_data} present a side-by-side comparison of the ground-truth trajectories of $[I_2, V_1, K]$ and those reconstructed by the Neural ODE. The close agreement further confirms the model’s ability to recover the correct dynamics from limited data, even when using an extended (quadratic) feature library---whereas SINDy fails to do so, even when provided with the full, densely sampled dataset.

\subsection{Examples where closure exists only after a coordinate transformation} \label{sec: neural_ode_non_closed_system}

Next, we investigate the performance of the Neural ODE framework on Example \eqref{ex:ex_2}, where the selected moments are $\bfx = [I_2, V_1, K, J]$. In this setting, the moment system is not closed unless an appropriate coordinate transformation is applied. A key question we explore here is: What qualitative and quantitative behaviors emerge when the Neural ODE is applied \textit{directly} to this unclosed system, without explicitly identifying the necessary coordinate transformation?

Following the setup in~\cite{10.3389/fphot.2024.1444993}, we adopt the nonlinear term $g(\rho, t) = \rho^2$ in Eq.~\eqref{eq:nls_harmonic} for Example~\ref{ex:ex_2}. To generate the moment trajectories, we numerically solve the PDE using the same initial condition as in Example~\ref{ex:ex_1}, given by Eq.~\eqref{ex:ex_1_IC_1}.

We first train a Neural ODE model using the full, uniformly sampled dataset and a degree-1 (linear) feature map $\mathcal{D}_{\deg = 1}(\mathbf{x}) = \begin{bmatrix} I_2, V_1,K, J \end{bmatrix}^\top$, without applying thresholding. The learned ODE system is:
\begin{align}
\label{ex:ex_2_lib_1_eqs}
\left\{
\begin{aligned}
    \frac{dI_2}{dt} &= 1.000V_1,\\
    \frac{dV_1}{dt} &= -2.000I_2 + 4.000K + 3.998J,\\
    \frac{dK}{dt} &= -0.005I_2-0.530V_1-0.003J+0.240J,\\
    \frac{dJ}{dt} &= 0.001I_2+0.030V_1+0.003J-0.161J.
\end{aligned}\right.
\end{align}

\begin{figure}[t]
    \centering
    \includegraphics[width=0.95\linewidth]{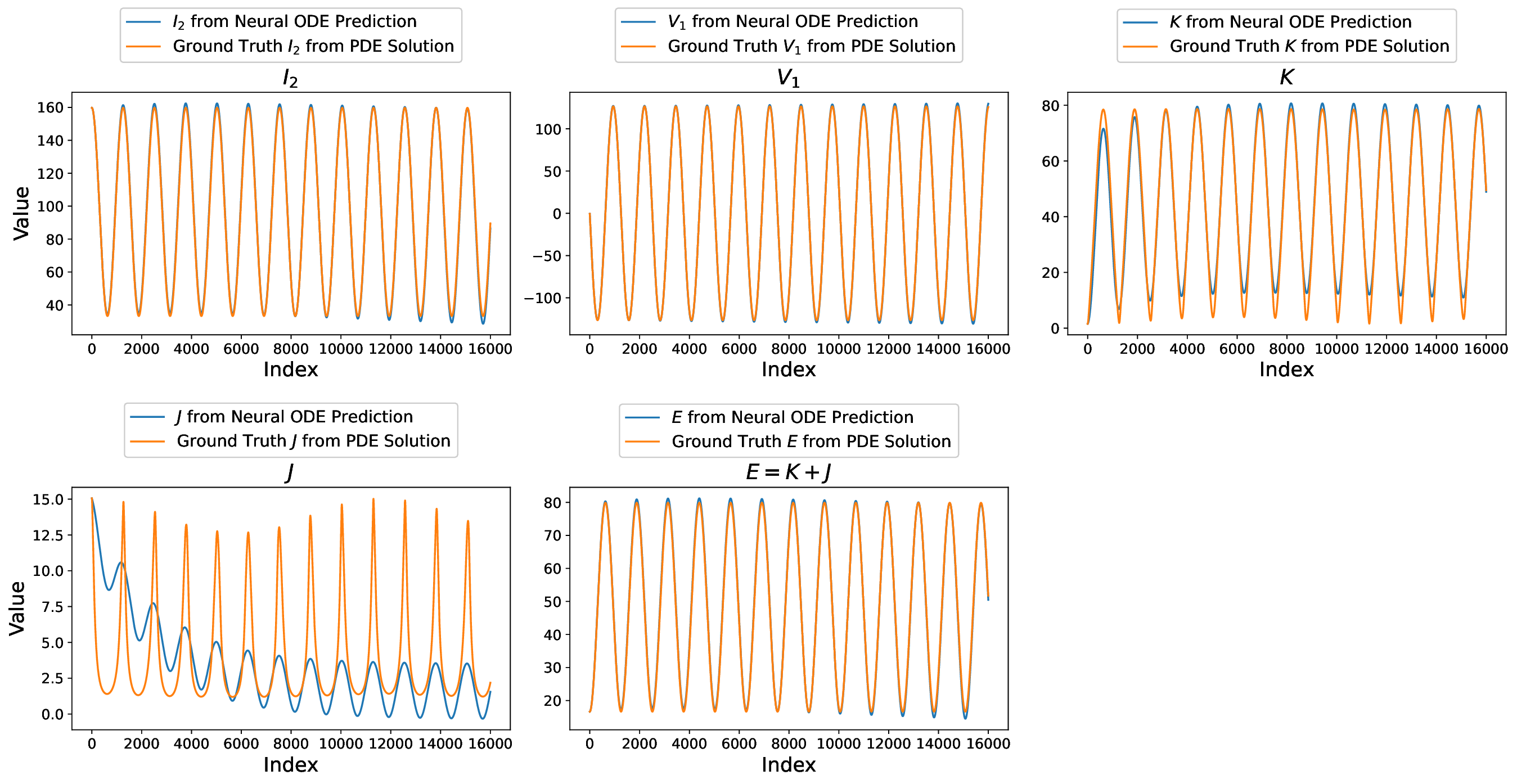}
    \caption{Comparison of ground-truth moment trajectories $[I_2, V_1, K, J, E=K+J]$ (from PDE simulations) with those reconstructed by Neural ODEs trained on full time-series data ($N = 16{,}000$ points), using a linear feature library. The model is trained solely on the selected moments $\bx = [I_2, V_1, K, J]$, without explicitly incorporating or being instructed to identify the closure transformation $E = K + J$.}
    \label{fig: IVKJE_full_data}
\end{figure}

The coefficients above are rounded to three decimal places. Figure~\ref{fig: IVKJE_full_data} (first four panels) compares the ground-truth trajectories of $[I_2, V_1, K, J]$, obtained from PDE simulation, to those predicted by integrating the learned Neural ODE system~\eqref{ex:ex_2_lib_1_eqs}. While the model accurately reproduces $I_2$ and $V_1$, it shows significant errors in $K$ and $J$, confirming that the system fails to recover the full dynamics in the original coordinate space $\mathbf{x} = [I_2, V_1, K, J]$---as expected, due to the absence of closure in these selected moments. These findings are quantitatively supported by the RMAE values reported in the first row of Table~\ref{tab:RMAE_ex2_IVKJ}.

\begin{table}[t]
\centering
\begin{tabular}{lcccccccc}
\toprule
\textbf{Variable} & $I_2$ & $V_1$ & $K$ & $J$ & $K+J$ & Overall (Orig.) \\
\midrule
\textbf{RMAE} for $[I_2, V_1, K, J]$ & 0.0098 & 0.0109 & 0.0318 & 0.1791 & N/A & 0.0127 \\
\midrule
\textbf{RMAE} for $[I_2, V_1, K+J]$ & 0.0098 & 0.0109 & N/A & N/A & 0.0094 & 0.0077 \\
\bottomrule
\end{tabular}
\caption{RMAE for each reconstructed moment variable and the overall error (aggregated over all moments), computed using a Neural ODE model trained on full time-series data ($N = 16{,}000$ points) with the linear feature map $\mathcal{D}_{\deg =1}(\mathbf{x})$. The model is trained on the selected moments $\bx = [I_2, V_1, K, J]$, without explicitly incorporating or discovering the coordinate transformation, $E = K + J$, required to close the system. The first row reports RMAE for the original moments $[I_2, V_1, K, J]$; the second row reports RMAE for $[I_2, V_1, K+J]$, obtained by summing the dynamics of $K$ and $J$. Notably, the RMAE for $E = K + J$ is significantly lower than for $K$ and $J$ individually.}
\label{tab:RMAE_ex2_IVKJ}
\end{table}

However, a closer examination of Eq.~\eqref{ex:ex_2_lib_1_eqs} shows that summing the equations for $K$ and $J$---effectively applying the known coordinate transformation needed for closure---yields an \textit{approximately} closed ODE system:
\begin{align}
 \label{ex:ex_2_transformed_lib_1}
\left\{
\begin{aligned}
    \frac{dI_2}{dt} &= 1.000V_1,\\
    \frac{dV_1}{dt} &= -2.000I_2 + 4.000(K+J) &&\quad- 0.002J,\\
    \frac{d\left(K+J\right)}{dt} &= -0.500V_1 &&\quad -0.004I_2 +0.079J.
\end{aligned}\right.
\end{align}

Despite small perturbative terms involving $I_2$ and $J$ in the dynamics of $E = K + J$, the transformed system remains remarkably consistent with the ground-truth moment equations in Eq.\eqref{eq:third_ME}, even though the model was not explicitly instructed to discover the coordinate transformation required for closure. This is further illustrated in the final panel of Figure~\ref{fig: IVKJE_full_data}, where the Neural ODE accurately reconstructs the trajectory of the combined variable $E$, despite failing to recover $K$ and $J$ individually. The RMAE values in the bottom row of Table~\ref{tab:RMAE_ex2_IVKJ} corroborate this finding, showing substantially lower error for $E=K+J$ compared to $K$ or $J$.

These results are slightly less precise than those in~\cite{10.3389/fphot.2024.1444993}, where the SINDy method introduces smaller extraneous terms compared to Eq.~\eqref{ex:ex_2_transformed_lib_1}. The Neural ODE framework also struggles to generalize when applied to richer feature spaces like $\mathcal{D}_{\deg = 1, 2}(\mathbf{x})$ or trained on sparse data. Moreover, thresholding offers limited benefit and can degrade performance when applied directly to an unclosed system.

These limitations highlight the need for Neural ODE models that can simultaneously learn necessary coordinate transformations for moment closure. We address this in the next section.

\subsection{Learning coordinate transformations for system closure via Neural ODE and Stiefel optimization} \label{sec: neural_ode_coordinate_transform}

As discussed in Sec. \ref{sec: neural_ode_non_closed_system}, the Neural ODE method struggles with systems that are unclosed in the original coordinate and cannot be expected to automatically identify a coordinate transformation for closure. To address this issue, we employ the method described in Sec.~\ref{sec:stiefel_neural_ode}, which combines Neural ODEs with coordinate transformation learning on the Stiefel manifold. Our results are primarily compared to those in~\cite{10.3389/fphot.2024.1444993}, where a linear feature map $\mathcal{D}_{\deg=1}(\bx)$ was used within the SINDy framework. For consistency and fair comparison, we also adopt a linear feature map. The general hyperparameter settings used in our method, as detailed in Algorithm~\ref{alg:inner_outer_optimization}, are: $\texttt{maxIter} = 1000$, $\texttt{Iter}{\texttt{scheduled}} = 800$, $\texttt{lr}{\texttt{initial}} = 10^{-3}$, and $\texttt{lr}{\texttt{scheduled}} = 10^{-4}$.

\noindent \textbf{Case 1: Initialization close to ground truth.} 
We begin by considering the case where the initial projection matrix $\mathbf{A}^{(0)}$ is chosen close to the ground-truth transformation $\mathbf{A}_{\text{gt}}$, which corresponds to taking the sum $K + J$ while leaving $I_2$ and $V_1$ unchanged, up to normalization to satisfy the Stiefel constraint. As described in Section~\ref{sec:neural_ode_method} and detailed in Algorithm~\ref{alg:inner_outer_optimization}, we promote sparsity in the learned dynamics by applying thresholding to the coefficient matrix $\mathbf{\Theta}^{(k)}$. The resulting learned coordinate transformation and dynamics matrices are:
\begin{align}
\label{eq:stifel_pred_sol}
    \widetilde{\mathbf{A}}_{\text{out}}
    \approx
    \begin{bmatrix}
        -1.000 &  0.000 &   -0.000 \\
        -0.000& -1.000&  0.000 \\
        0.000& -0.000& -0.982\\
        0.000 & 0.000 &-0.188
    \end{bmatrix}, \quad
    \widetilde{\mathbf{\Theta}}_{\text{out}}
    \approx
    \begin{bmatrix}
        0& 0.792&   0 \\
        -2.526&   0& 2.533\\
        0&  -0.790&0
    \end{bmatrix}.
\end{align}
The coefficients are rounded to three decimal places. For reference, the ground-truth coordinate transformation $\mathbf{A}_{\text{gt}}$ (normalized to satisfy the Stiefel manifold constraint), along with the corresponding dynamics matrix $\mathbf{\Theta}_{\text{gt}}$ derived from Eq.~\eqref{eq:third_ME}, are given below. Here, $C_{I_2}$, $C_{V_1}$, $C_K$, and $C_J$ denote the normalization constants:
\begin{align}
\label{eq:stifel_gt_sol}
\mathbf{A}_{\text{gt}} &=
\begin{bmatrix}
1 & 0 & 0 \\
0 & 1 & 0 \\
0 & 0 & \tfrac{C_K}{C_J} \big/ \sqrt{\left(1 + \tfrac{C_K}{C_J} \right)^2} \\
0 & 0 & \tfrac{1}{\sqrt{\left(1 + \tfrac{C_K}{C_J} \right)^2}}
\end{bmatrix}
\approx
\begin{bmatrix}
1 & 0 & 0 \\
0 & 1 & 0 \\
0 & 0 & 0.982 \\
0 & 0 & 0.188
\end{bmatrix}, \\[1em]
\mathbf{\Theta}_{\text{gt}} &=
\begin{bmatrix}
0 & \tfrac{C_{V_1}}{C_{I_2}} & 0 \\
-2\tfrac{C_{I_2}}{C_{V_1}} & 0 & \tfrac{4\sqrt{C_K^2 + C_J^2}}{C_{V_1}} \\
0 & -\tfrac{1}{2} \tfrac{C_{V_1}}{\sqrt{C_K^2 + C_J^2}} & 0
\end{bmatrix}
\approx
\begin{bmatrix}
0 & 0.792 & 0 \\
-2.526 & 0 & 2.533 \\
0 & -0.790 & 0
\end{bmatrix}.
\end{align}
As shown in Eq.\eqref{eq:stifel_gt_sol}, when the initialization is close to the ground-truth matrices $(\mathbf{A}_{\text{gt}}, \mathbf{\Theta}_{\text{gt}})$, the learned solution $(\widetilde{\mathbf{A}}_{\text{out}}, \widetilde{\mathbf{\Theta}}_{\text{out}})$ in Eq.\eqref{eq:stifel_pred_sol} closely matches the true system. This demonstrates that, with proper initialization and sparsity-promoting optimization via thresholding, the proposed framework not only recovers the correct coordinate transformation required for closure but also yields a sparse and interpretable representation of the governing dynamics.

\noindent \textbf{Case 2: Random initialization}. 
We next consider the scenario where the projection matrix $\mathbf{A}^{(0)}$ in Eq.~\eqref{eq:optimization_inner_outer} is initialized randomly. As in the previous case, we use the linear feature library $\mathcal{D}_{\deg=1}(\bx)$. The algorithm then returns:
\begin{align}
\label{eq:stifel_pred_sol_trans}
\widetilde{\mathbf{A}}_{\text{out}}
\approx
\begin{bmatrix}
-0.449 & -0.622 &  \phantom{-}0.642 \\
-0.401 & -0.502 & -0.766 \\
 \phantom{-}0.784 & -0.591 & -0.023 \\
 \phantom{-}0.150 & -0.113 & -0.004
\end{bmatrix}, \quad
\widetilde{\boldsymbol{\Theta}}_{\text{out}}
\approx
\begin{bmatrix}
-0.870 &  \phantom{-}0.476 &  \phantom{-}1.430 \\
-1.578 & -0.015 &  \phantom{-}0.858 \\
-2.631 & -0.301 &  \phantom{-}0.885
\end{bmatrix}.
\end{align} 
At first glance, the predicted solution in Eq.~\eqref{eq:stifel_pred_sol} appears to differ from the ground truth presented in Eq.\eqref{eq:stifel_gt_sol}. However, as noted in Remark 2 of~\cite{10.3389/fphot.2024.1444993}, even under the Stiefel manifold constraint $\mathbf{A}^\top \mathbf{A} = \mathbf{I}_{3 \times 3}$, the solution to Eq.~\eqref{eq:optimization_inner_outer} is not unique. In particular, if $(\mathbf{A}, \mathbf{\Theta})$ is a solution, then for any orthogonal matrix $\mathbf{O} \in \mathrm{O}(3)$, the pair
\begin{align}
\widetilde{\mathbf{A}}^* = \mathbf{A}^* \mathbf{O}, \quad
\widetilde{\mathbf{\Theta}}^* = \mathbf{O}^\top \mathbf{\Theta}^* \mathbf{O}
\end{align}
is also a valid solution of problem~\eqref{eq:optimization_inner_outer} that yields the same reconstruction error.

To verify whether the solution in Eq.~\eqref{eq:stifel_pred_sol_trans} is equivalent to the ground-truth solution in Eq.~\eqref{eq:stifel_gt_sol}, we apply an additional change of coordinates using:
\begin{align}
    \mathbf{O} =
    \begin{bmatrix}
        -0.449 & -0.401 & \phantom{-}0.799 \\
        -0.622 & -0.502 & -0.601 \\
         \phantom{-}0.642 & -0.766 & -0.024
    \end{bmatrix} \in \text{O}(3),
\end{align}
then we have:

\begin{align}
\label{eq:stifel_pred_sol_lasso_transform}
\widetilde{\mathbf{A}}_{\text{out}} \, \widetilde{\mathbf{O}}
\approx
\begin{bmatrix}
1.000 & -0.000 & \phantom{-}0.000 \\
-0.000 & \phantom{-}1.000 & \phantom{-}0.000 \\
-0.000 & \phantom{-}0.000 & \phantom{-}0.982 \\
-0.000 & -0.000 & \phantom{-}0.188
\end{bmatrix}
\approx \mathbf{A}_{\text{gt}}, \quad
\widetilde{\mathbf{O}}^\top \, \widetilde{\mathbf{\Theta}}_{\text{out}} \, \widetilde{\mathbf{O}}
\approx
\begin{bmatrix}
\phantom{-}0.000 & \phantom{-}0.792 & -0.000 \\
-2.526 & -0.000 & \phantom{-}2.533 \\
-0.000 & -0.790 & -0.000
\end{bmatrix}
\approx \mathbf{\Theta}_{\text{gt}}.
\end{align}

Hence, by Remark 2 in~\cite{10.3389/fphot.2024.1444993}, the recovered solution 
$(\widetilde{\mathbf{A}}_{\text{out}} \, \widetilde{\mathbf{O}}, \widetilde{\mathbf{O}}^\top \, \widetilde{\mathbf{\Theta}}_{\text{out}} \, \widetilde{\mathbf{O}})$ is equivalent to the ground truth $(\mathbf{A}_{\text{gt}}, \mathbf{\Theta}_{\text{gt}})$ up to an orthogonal transformation. This confirms that our method successfully identifies a valid coordinate transformation $\mathbf{A}_{\text{out}}$ that closes the moment system.

However, under random initialization, the learned dynamics matrix $\mathbf{\Theta}_{\text{out}}$ is not sparse, as no regularization penalizes model complexity. In this setting, thresholding is less effective, as the solution is non-unique over the Stiefel manifold. The algorithm may converge to a nonsparse, yet functionally equivalent, solution that deviates from $\mathbf{A}_{\text{gt}}$. {Nonetheless, post-processing of the learned $\widetilde{\mathbf{A}}_{\text{out}}$ and $\widetilde{\mathbf{\Theta}}_{\text{out}}$---such as identifying an orthogonal transformation $\widetilde{\mathbf{O}} \in \text{O}(3)$ to sparsify the dictionary coefficients $\widetilde{\mathbf{\Theta}}_{\text{out}}$, as in Eq.~\eqref{eq:stifel_pred_sol_lasso_transform}---can be performed to enable post hoc learning of a compact and interpretable closed-form moment dynamics.
}

We also evaluated the framework under significant data sparsity, using only 4,000 randomly selected samples from 16,000. Even under this constraint, the method accurately identified both the coordinate transformation and the corresponding closed ODE system, demonstrating its robustness and efficiency in data-limited settings.

\subsection{Neural ODE approximation of the non-closed system} \label{sec: neural_ode_FK}
Finally, we apply the Neural ODE framework to systems without exact analytical closures, focusing on the example introduced in Example~\ref{ex:ex_3}. Our objective is to compare the manually derived approximate moment system from~\cite{BELMONTEBEITIA20143267} with a data-driven Neural ODE model and examine whether improved or alternative moment dynamics can be identified. Model performance is evaluated through extrapolation: the Neural ODE is trained on the first 200 of 300 time points, and predictions are made for the remaining 100. Accuracy is quantified using the RMAE on the test set. We also assess performance under data sparsity by randomly selecting 100 training samples from the first 200 time points, followed by the same extrapolation-based evaluation.

\begin{figure}[hbt!]
    \centering
    \includegraphics[width=1\linewidth]{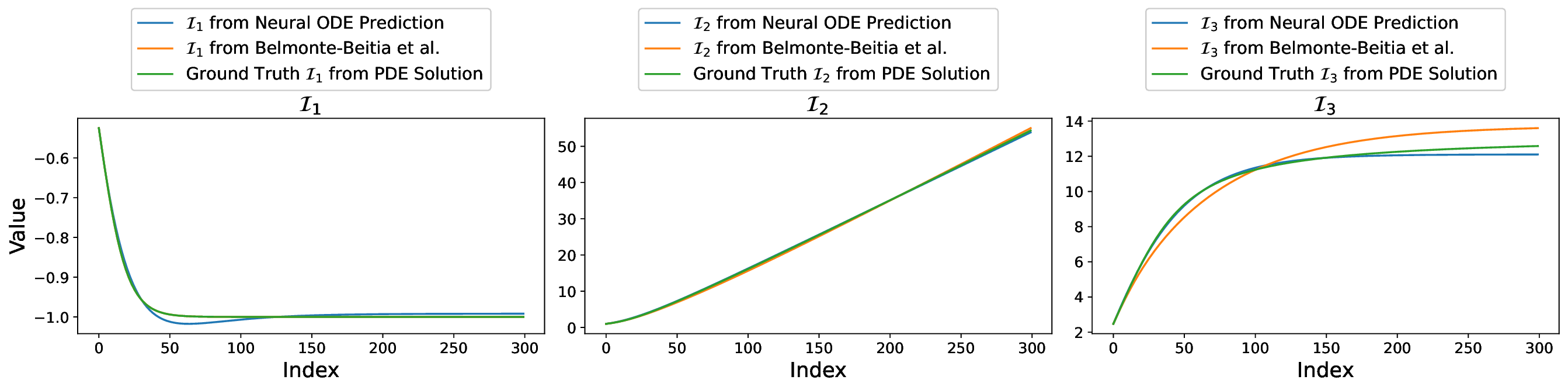}
    \caption{Reconstructed trajectories of $[\mathcal{I}_1, \mathcal{I}_2, \mathcal{I}_3]$ over the full time horizon, based on a Neural ODE model trained on the first 200 of 300 time points using a quadratic feature library $\mathcal{D}_{\deg \le 2}(\mathbf{x})$. Results are compared with the analytical model from Belmonte-Beitia et al.~\cite{BELMONTEBEITIA20143267} and the ground-truth moment trajectories obtained from PDE simulations.}
    \label{fig: ex4_quadratic_200_data}
\end{figure}

We consider a polynomial feature map that includes terms up to second degree $\mathcal{D}_{\deg \le 2}(\mathbf{X}) = \begin{bmatrix} 1,\ \mathcal{I}_1,\ \mathcal{I}_2,\ \mathcal{I}_3,\ \mathcal{I}_1^2,\ \mathcal{I}_2^2,\ \mathcal{I}_3^2,\ \mathcal{I}_1 \mathcal{I}_2,\ \mathcal{I}_1 \mathcal{I}_3,\ \mathcal{I}_2 \mathcal{I}_3 \end{bmatrix}^\top$. To promote sparsity and interpretability, we augment the Neural ODE training by including an $\ell_1$ regularization term on the coefficients in the loss function (Eq.~\eqref{eq:neural_ode_loss_batch}) and apply iterative thresholding every 1,000 iterations. This encourages compact representations by removing redundant terms. The resulting learned system, with coefficients rounded to three decimal places, is:
\begin{align}
\label{eq:neural_ode_ex4_learned}
\left\{
\begin{aligned}
\frac{d\mathcal{I}_1}{dt} &= -0.338 + 0.254\,\mathcal{I}_1^2 - 0.007\,\mathcal{I}_1\mathcal{I}_3, \\
\frac{d\mathcal{I}_2}{dt} &= -0.356\,\mathcal{I}_1 + 0.083\,\mathcal{I}_3 + 0.544\,\mathcal{I}_1^2, \\
\frac{d\mathcal{I}_3}{dt} &= 1.890 - 0.499\,\mathcal{I}_1 - 0.043\,\mathcal{I}_3 - 0.007\,\mathcal{I}_3^2 + 0.068\,\mathcal{I}_1\mathcal{I}_3.
\end{aligned}
\right.
\end{align}
In Figure~\ref{fig: ex4_quadratic_200_data}, we compare the reconstructed trajectories of the moments $[\mathcal{I}_1, \mathcal{I}_2, \mathcal{I}_3]$---obtained by integrating the Neural ODE trained on the first 200 time points---with those from the analytically derived system proposed by Belmonte-Beitia et al.~\cite{BELMONTEBEITIA20143267} and the ground-truth trajectories from PDE simulations. The Neural ODE achieves comparable accuracy to the analytical model on $\mathcal{I}_2$, performs slightly worse on $\mathcal{I}_1$, and substantially outperforms it on $\mathcal{I}_3$.

\begin{table}[h!]
\centering
\begin{tabular}{lccccc}
\toprule
\textbf{Method} & $\mathcal{I}_1$ & $\mathcal{I}_2$ & $\mathcal{I}_3$ & Overall (Norm.) & Overall (Orig.) \\
\midrule
\textbf{Neural ODE (200 full data)} & 0.0078 & 0.0045 & 0.0280 & 0.0134 & 0.0037 \\
\textbf{Neural ODE (100 sparse data)} & 0.0076 & 0.0103 & 0.0277 & 0.0152 & 0.0056 \\
\textbf{Belmonte-Beitia et al.} & 0.0000 & 0.0055 & 0.0784 & 0.0280 & 0.0079 \\
\bottomrule
\end{tabular}
\caption{
RMAE for each reconstructed moment variable and overall errors. The Neural ODE is trained either on the full set of the first 200 data points or on its sparse subset of 100 randomly selected samples, using a quadratic feature map $\mathcal{D}_{\deg \le 2}(\mathbf{X})$. The RMAE is computed on extrapolation data---100 future time points not seen during training. ``Norm.'' indicates errors computed in normalized space; ``Orig.'' indicates errors computed after rescaling to physical units.
}
\label{tab:rmae_ex4}
\end{table}

We then quantitatively evaluate and compare the predictive accuracy of both models over the extrapolation range (time steps 201–300). The corresponding RMAE values are summarized in Table~\ref{tab:rmae_ex4}. These results highlight the superior extrapolation performance of the Neural ODE model relative to the analytical system derived by Belmonte-Beitia et al. While the analytical model yields a slightly lower error for $\mathcal{I}_1$, this advantage is minor due to the small magnitude of $\mathcal{I}_1$ (typically between –0.5 and –1), which contributes less to the overall error. In contrast, the Neural ODE model achieves substantially lower total error in both normalized and original physical coordinates. Remarkably, even under a sparse data regime---where only 100 training points are randomly selected from the first 200---our Neural ODE model maintains strong predictive performance. The resulting overall RMAE (0.0152 normalized, 0.0056 original) remains close to that of the full 200-sample model (0.0134 normalized, 0.0037 original), underscoring the method’s robustness to data scarcity. Additional results and implementation details are provided in the appendix. These results demonstrate that the proposed data-driven approach not only effectively captures the approximate dynamics of non-closed systems but also generalizes well under data scarcity.

In addition to the polynomial feature map of degree $ \leq 2 $, we construct an extended feature library that includes the fractional-degree terms found in the manually derived system of Eq.~\eqref{eq:third_ME_2}. These terms include, for example, $ \sqrt{\mathcal{I}_1} $ (degree $ 1/2 $) and $ \sqrt{\mathcal{I}_1}\,\mathcal{I}_2 $ (degree $ 3/2 $), along with their mixed products. 
This extension is motivated
by the particular form of the
ODEs of Eqs.~(\ref{eq:third_ME_2}).
The resulting feature map is
\[
\mathcal{D}_{\text{extended}}(\mathbf{X}) =
\left[
\begin{array}{cccccccccc}
1 & \mathcal{I}_1 & \mathcal{I}_2 & \mathcal{I}_3 &
\sqrt{\mathcal{I}_1} & \sqrt{\mathcal{I}_2} & \sqrt{\mathcal{I}_3} &
\mathcal{I}_1^{2} & \mathcal{I}_2^{2} & \mathcal{I}_3^{2} \\
\mathcal{I}_1 \mathcal{I}_2 & \mathcal{I}_1 \mathcal{I}_3 & \mathcal{I}_2 \mathcal{I}_3 &
\sqrt{\mathcal{I}_1}\mathcal{I}_2 & \sqrt{\mathcal{I}_1}\mathcal{I}_3 &
\sqrt{\mathcal{I}_2}\mathcal{I}_1 & \sqrt{\mathcal{I}_2}\mathcal{I}_3 &
\sqrt{\mathcal{I}_3}\mathcal{I}_1 & \sqrt{\mathcal{I}_3}\mathcal{I}_2
\end{array}
\right]^{\!\top}
\]

{This 19‑term library spans degrees 0, 1/2, 1, 3/2, and 2, enabling the Neural ODE to capture both integer‑and fractional‑order dynamics. Interested readers are referred to the appendix for the full learned moment system and corresponding extrapolation results. {While this enriched representation enhances the expressiveness of the learned dynamics, it does not lead to a substantial improvement in extrapolation accuracy compared to the simpler quadratic feature map. This highlights a key insight: although increasing the feature space can capture more complex interactions, it does not necessarily yield better predictive performance, underscoring a fundamental trade-off between model expressiveness and generalization.}}

\section{Conclusions and future works}
\label{sec:conclusion}
In the present work we proposed a data-driven framework based on Neural ODEs for learning reduced-order dynamical systems that describe or approximate the evolution of selected moment variables in PDEs. The goal is to extract closed-form, low-dimensional representations that are both parsimonious and accurate.

We first showed that when the moment system is analytically closed, Neural ODEs can accurately recover the governing equations---even under sparse and/or irregular sampling---using a prescribed (potentially overcomplete) feature map. In contrast, SINDy-based methods often fail in such settings due to their reliance on time-derivative estimation.

For unclosed systems, directly applying Neural ODEs often leads to less
accurate results. To address this, we developed an algorithm that jointly learns a (linear) coordinate transformation and the governing equations in the transformed space. This approach identifies transformations that approximately close the system, even under data sparsity, and yields sparse, interpretable dynamics when combined with thresholding and regularization.

We further validated the method on the Fisher–KPP equation, whose moment system lacks an exact analytical closure. The learned Neural ODE model accurately captured key dynamics and outperformed the manually derived approximation
in a suitably defined 
quantitative sense, even under sparse-data conditions.

These results underscore the potential of Neural ODEs as a flexible, data-driven alternative to classical and other data-driven moment closure techniques. In future work, we believe
that the promising results herein
suggest extensions to noisy and multi-fidelity data as well as ones stemming
from higher dimensional PDEs, generalizations to non-polynomial or structured feature spaces, and the discovery of nonlinear coordinate transformations needed for closure.
The latter may be an especially 
important direction where the naive
or physically inspired moments do
not appear to close. These 
directions are currently under study
and will be reported in future
publications.

\section*{Acknowledgements}
This material is based upon work supported by the U.S. National Science Foundation under awards DMS-2052525, DMS-2140982, DMS-2244976 (WZ), PHY-2110030, PHY-2408988, and DMS-2204702 (PGK). WZ is also partially supported by the AFOSR grant FA9550-25-1-0079.

\newpage 

\newpage

\bibliographystyle{unsrt}
\bibliography{main}

\newpage
\appendix
\onecolumn

\section*{Appendix}
In this appendix, we provide additional experimental results that complement the ones presented in the main text.

\appendix

\section{Neural ODE Approximation of the Non-Closed System}

\subsection{Sparse Data: Quadratic Feature Map with 100 Training Samples}

To further evaluate the performance of our approach under data scarcity, we trained the Neural ODE using only 100 randomly selected samples from the first 200 time steps. We maintained the same quadratic feature map $\mathcal{D}_{\deg \le 2}(\mathbf{x})$ and applied $\ell_1$ regularization with iterative thresholding. The learned moment dynamics, after rounding coefficients to three decimal places, are given by:

\begin{align}
\label{eq:neural_ode_ex4_learned_new}
\left\{
\begin{aligned}
\frac{d\mathcal{I}_1}{dt} &= -0.316 + 0.237\,\mathcal{I}_1^2 - 0.007\,\mathcal{I}_1\mathcal{I}_3, \\
\frac{d\mathcal{I}_2}{dt} &= 0.060 - 0.521\,\mathcal{I}_1 + 0.042\,\mathcal{I}_3 + 0.206\,\mathcal{I}_1^2 - 0.054\,\mathcal{I}_1\mathcal{I}_3, \\
\frac{d\mathcal{I}_3}{dt} &= 1.884 - 0.470\,\mathcal{I}_1 - 0.027\,\mathcal{I}_3 - 0.008\,\mathcal{I}_3^2 + 0.068\,\mathcal{I}_1\mathcal{I}_3.
\end{aligned}
\right.
\end{align}

\begin{figure}[hbt!]
    \centering
    \includegraphics[width=1\linewidth]{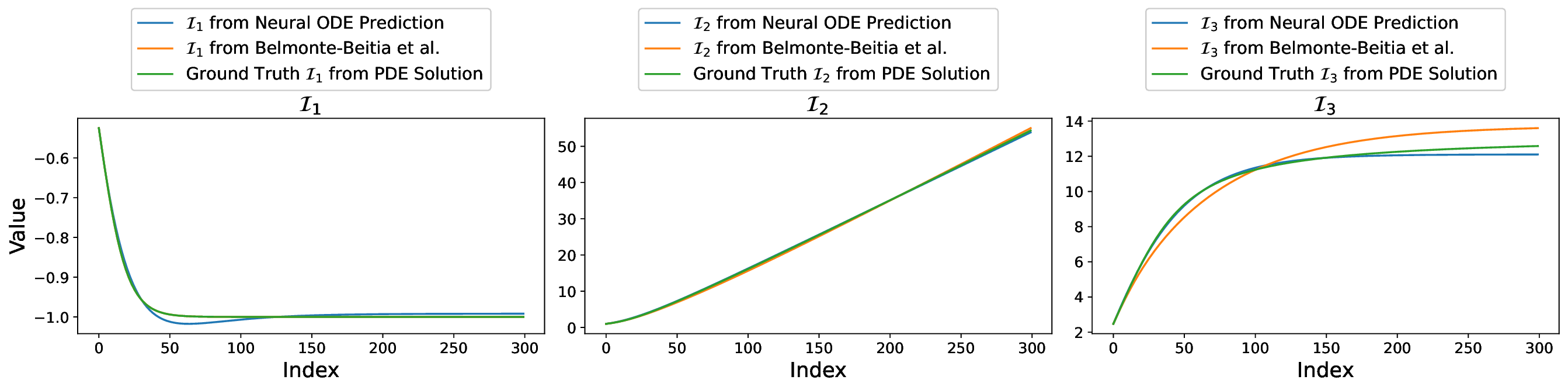}
    \caption{
Reconstructed trajectories of $[\mathcal{I}_1, \mathcal{I}_2, \mathcal{I}_3]$ obtained by integrating the Neural ODE learned from 100 randomly selected training points out of the first 200 time steps, using a quadratic feature map $\mathcal{D}_{\deg \le 2}(\mathbf{x})$. The results are compared against the analytical model proposed by Belmonte-Beitia et al.~\cite{BELMONTEBEITIA20143267} and the ground truth solution of the PDE.}
    \label{fig: ex4_quadratic_200_100_sparse_data}
\end{figure}

Figure~\ref{fig: ex4_quadratic_200_100_sparse_data} shows the reconstructed trajectories of the moments $[\mathcal{I}_1, \mathcal{I}_2, \mathcal{I}_3]$, obtained by integrating the Neural ODE trained on the 100 randomly selected samples. The results are compared with the analytical model proposed by Belmonte-Beitia et al.~\cite{BELMONTEBEITIA20143267} and the ground-truth PDE simulations. Despite the reduced training data, the Neural ODE achieves comparable accuracy to the analytical model on $\mathcal{I}_2$, performs slightly worse on $\mathcal{I}_1$, and substantially outperforms it on $\mathcal{I}_3$.

Despite the reduced training data, the model demonstrates excellent generalization in extrapolation. As shown in Table~\ref{tab:rmae_ex4_sparse}, the resulting system achieves an overall normalized RMAE of 0.0152 and a physical RMAE of 0.0056—values that are close to those obtained using the full 200-sample training set in Sec.~\ref{sec: neural_ode_FK}. These findings further confirm the robustness of our Neural ODE framework in learning compact and interpretable approximations of non-closed systems, even when only limited and irregular data are available.

\begin{table}[h!]
\centering
\begin{tabular}{lccccc}
\toprule
\textbf{Method} & $\mathcal{I}_1$ & $\mathcal{I}_2$ & $\mathcal{I}_3$ & Overall (Norm.) & Overall (Orig.) \\
\midrule
\textbf{Neural ODE} & 0.0076 & 0.0103 & 0.0277 & 0.0152 & 0.0056 \\
\textbf{Belmonte-Beitia et al.} & 0.0000 & 0.0055 & 0.0784 & 0.0280 & 0.0079 \\
\bottomrule
\end{tabular}
\caption{
RMAE for each reconstructed moment variable and overall extrapolation error over 100 unseen time steps. The Neural ODE is trained on 100 randomly selected samples from the first 200 time steps using a quadratic feature map $\mathcal{D}_{\deg \le 2}(\mathbf{x})$. Results are compared with the analytical model of Belmonte-Beitia et al.~\cite{BELMONTEBEITIA20143267}. “Norm.” refers to errors computed in normalized space; “Orig.” refers to errors rescaled to physical units.
}
\label{tab:rmae_ex4_sparse}
\end{table}

\subsection{Beyond Quadratics: Extended Feature Map with 200 Training Samples} 

To enable a direct comparison with the manually derived system, we extended the feature map to include fractional-degree terms such as square roots and their mixed products. While this enriched library allows the Neural ODE to represent a broader class of nonlinear interactions, the resulting model does not outperform the one constructed using the simpler quadratic feature map in terms of overall extrapolation accuracy. 

\begin{align}
\label{eq:neural_ode_ex4_learned_extended}
\left\{
\begin{aligned}
\frac{d\mathcal{I}_1}{dt} &= -0.2920 + 0.0231\,\mathcal{I}_1 + 0.0683\,\sqrt{\mathcal{I}_1} - 0.0234\,\mathcal{I}_2^2 + 0.1673\,\sqrt{\mathcal{I}_1}\,\mathcal{I}_2, \\
\frac{d\mathcal{I}_2}{dt} &= 0.1375 - 0.0973\,\mathcal{I}_1 + 0.0344\,\mathcal{I}_3 - 0.0756\,\sqrt{\mathcal{I}_1} + 0.0141\,\mathcal{I}_1^2 + 0.6660\,\sqrt{\mathcal{I}_1}\,\mathcal{I}_2 - 0.0595\,\sqrt{\mathcal{I}_3}\,\mathcal{I}_1, \\
\frac{d\mathcal{I}_3}{dt} &= 1.8800 - 0.7036\,\mathcal{I}_1 - 0.0419\,\mathcal{I}_3 + 0.3109\,\sqrt{\mathcal{I}_1} - 0.0178\,\sqrt{\mathcal{I}_3} + 0.0456\,\mathcal{I}_2^2 \\
&\quad - 0.0543\,\mathcal{I}_1\mathcal{I}_3 - 0.0928\,\sqrt{\mathcal{I}_1}\,\mathcal{I}_2 + 0.0594\,\sqrt{\mathcal{I}_3}\,\mathcal{I}_1.
\end{aligned}
\right.
\end{align}

\begin{figure}[hbt!]
    \centering
    \includegraphics[width=1\linewidth]{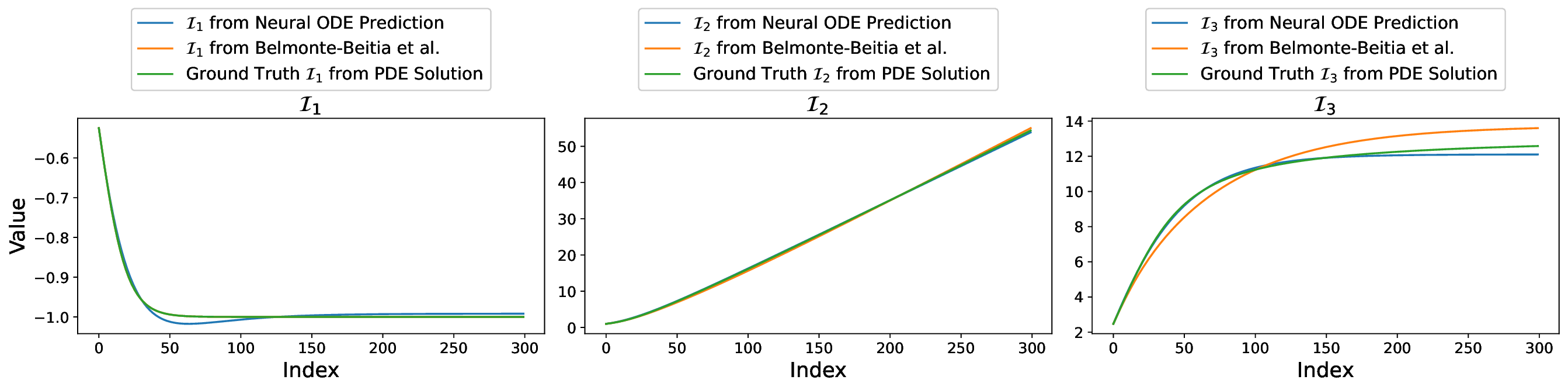}
    \caption{Reconstructed trajectories of  $[\mathcal{I}_1, \mathcal{I}_2, \mathcal{I}_3]$ over the full time horizon, based on a Neural ODE model trained on 200 out of 300 time points using the extended feature library $\mathcal{D}_{\text{extended}}(\mathbf{x})$. The results are compared against the analytical model from Belmonte-Beitia et al.~\cite{BELMONTEBEITIA20143267} and the ground truth solution of the PDE.}
    \label{fig: ex4_qubic_200_data}
\end{figure}

Figure~\ref{fig: ex4_qubic_200_data} presents the reconstructed trajectories of the moments $[\mathcal{I}_1, \mathcal{I}_2, \mathcal{I}_3]$ obtained by integrating the Neural ODE trained on the first 200 time steps. The results are compared against the analytical model proposed by Belmonte-Beitia et al.~\cite{BELMONTEBEITIA20143267} and the ground-truth PDE simulations. As shown in Table~\ref{tab:rmae_ex4_qubic}, the extended model exhibits slightly higher errors in reconstructing $\mathcal{I}_1$ and $\mathcal{I}_2$, leading to a higher overall RMAE. This outcome suggests that increasing the complexity of the feature space does not necessarily enhance predictive performance and may, in some cases, introduce overfitting or hinder optimization. Nonetheless, the extended Neural ODE still clearly outperforms the manually derived model from~\cite{BELMONTEBEITIA20143267}, particularly in the reconstruction of $\mathcal{I}_3$, highlighting the robustness and adaptability of data-driven modeling even in analytically non-closed systems.

\begin{table}[hbt!]
\centering
\begin{tabular}{lccccc}
\toprule
\textbf{Method} & $\mathcal{I}_1$ & $\mathcal{I}_2$ & $\mathcal{I}_3$ & Overall (Norm.) & Overall (Orig.) \\
\midrule
\textbf{Neural ODE} & 0.0092 & 0.0139 & 0.0115 & 0.0116 & 0.0056 \\
\textbf{Belmonte-Beitia et al.} & 0.0000 & 0.0055 & 0.0784 & 0.0280 & 0.0079 \\
\bottomrule
\end{tabular}
\caption{
RMAE for each reconstructed moment variable and overall errors computed on extrapolation data (100 time points) using a extended feature map $\mathcal{D}_{\text{extended}}(\mathbf{x})$. “Norm.” indicates errors computed in normalized space; “Orig.” indicates errors computed after rescaling to physical units.
}
\label{tab:rmae_ex4_qubic}
\end{table}

\end{document}